\documentclass[times, twoside, watermark]{zHenriquesLab-StyleBioRxiv}

\usepackage{flushend}

\leadauthor{Zhou} 

\begin{document}

\title{Chance, long tails, and inference: \\
a non-Gaussian, Bayesian theory of \\
vocal learning in songbirds}

\author[a]{Baohua Zhou}
\author[a,b]{David Hofmann}
\author[a,b]{Itai Pinkoviezky}
\author[c]{Samuel J.\ Sober}
\author[a,b,c,1]{Ilya Nemenman}

\affil[a]{Department of Physics, Emory University, Atlanta, GA 30322}
\affil[b]{Initiative in Theory and Modeling of Living Systems, Emory University, Atlanta, GA 30322}
\affil[c]{Department of Biology, Emory University, Atlanta, GA 30322}

\maketitle

\begin{abstract}
\noindent Traditional theories of sensorimotor learning posit that animals use sensory error signals to find the optimal motor command in the face of Gaussian sensory and motor noise. However, most such theories cannot explain common behavioral observations, for example that smaller sensory errors are more readily corrected than larger errors and that large abrupt (but not gradually introduced) errors lead to weak learning. Here we propose a new theory of sensorimotor learning that explains these observations. The theory posits that the animal learns an entire probability distribution of motor commands rather than trying to arrive at a single optimal command, and that learning arises via Bayesian inference when new sensory information becomes available. We test this theory using data from a songbird, the Bengalese finch, that is adapting the pitch (fundamental frequency) of its song following perturbations of auditory feedback using miniature headphones. We observe the distribution of the sung pitches to have long, non-Gaussian tails, which, within our theory, explains the observed dynamics of learning. Further, the theory makes surprising predictions about the dynamics of the shape of the pitch distribution, which we confirm experimentally.
\end {abstract}


\begin{corrauthor}
E-mail: ilya.nemenman@emory.edu
\end{corrauthor}

\section*{Introduction}
Learned behaviors --- reaching for an object, walking, talking, and hundreds of others --- allow the organism to interact with the ever-changing surrounding world. To learn and execute skilled behaviors, it is vital for such behaviors to fluctuate from iteration to iteration. Such variability is not limited to inevitable biological noise~\cite{shadmehr2010,mcdonnell2011}, but rather a significant part of it is controlled by animals themselves and is used for exploration during learning~\cite{neuringer2002,kao2005}. Furthermore, learned behaviors rely heavily on sensory feedback. The feedback is needed, first, to guide the initial acquisition of the behaviors, and then to maintain the needed motor output in the face of changes in the motor periphery and fluctuations in the environment. Within such sensorimotor feedback loops, the brain computes how to use the inherently noisy sensory signals to change patterns of activation of inherently noisy muscles to produce the desired behavior. This transformation from sensory feedback to motor output is both robust and flexible, as demonstrated in many species in which systematic perturbations of the feedback dramatically reshape behaviors~\cite{linkenhoker2002,knudsen2002,smith2006,sober2009,shadmehr2010}.

Since many complex behaviors are characterized by both tightly controlled motor variability and by robust sensorimotor learning, we propose that, during learning, the brain controls the {\em distribution of behaviors}. In contrast, most prior theories of animal learning have assumed that there is a single optimal motor command that the animal tries to produce, and that, after learning, deviations from the optimal behavior result from the unavoidable (Gaussian) downstream motor noise. Such prior models include the classic Rescorla-Wagner (RW) model~\cite{rw72}, as well as more modern approaches belonging to the family of reinforcement learning~\cite{joel2002, sutton2012, lak2016}, Kalman filters~\cite{kording2007,wolpert2007}, or dynamical Bayesian filter models~\cite{gallistel2001,gershman2015}. Such theories have addressed many important experimental questions, such as evaluating the optimality of the learning process~\cite{donchin2001,kording2007,van_beers2009,wei2009,beck2012}, accounting for multiple temporal scales in learning~\cite{smith2006,kording2007,wei2010,kelly2014}, identifying the complexity of behaviors that can be learned~\cite{genewein2015}, and pointing out how the necessary computations could be performed using networks of spiking neurons~\cite{shadmehr2005,dayan2008,fischer2011,neymotin2013,schultz2015,lak2016}.

However, despite these successes, most prior models that assume that the brain aims to achieve a single optimal output have been unable to explain some commonly observed experimental results. For example, since such theories assume that errors between the target and the realized behavior drive changes in future motor commands, they typically predict large behavioral changes in response to large errors. In contrast, experiments in multiple species report a decrease in both the speed and the magnitude of learning with an increase in the experienced sensory error~\cite{knudsen2002,robinson2003,wei2009,sober2012}. One can rescue traditional theories  by allowing the animal to {\em reject} large errors as ``irrelevant''---unlikely to have come from its own actions~\cite{wei2009, hahnloser2017}. However, such rejection models have not yet explained why the same animals that cannot compensate for large errors can correct for even larger ones, as long as their magnitude grows gradually with time~\cite{knudsen2002,sober2012}.

\begin{SCfigure*}[0.5385][t]
\centering
\includegraphics[width=0.65\textwidth]{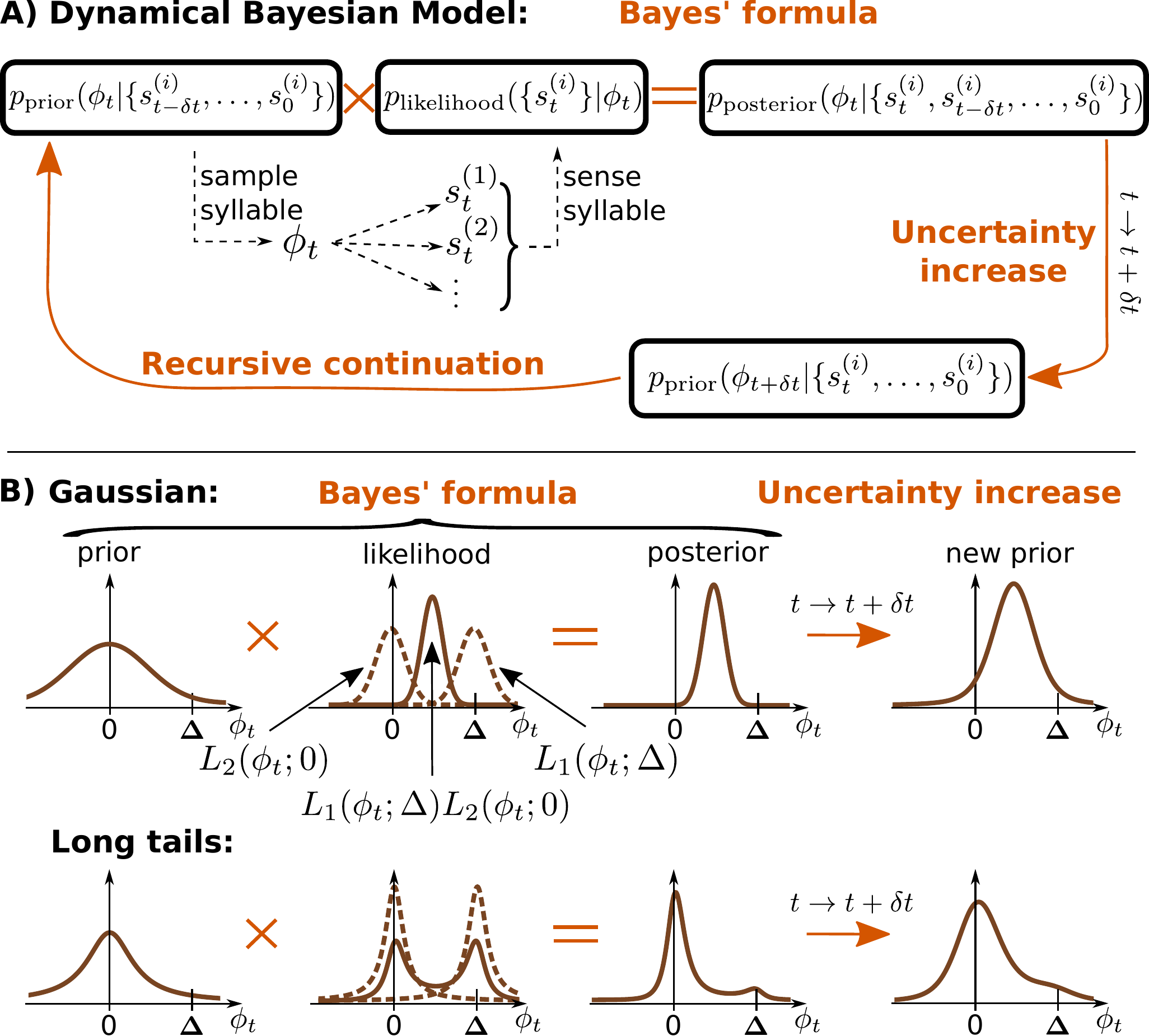}
\caption{The dynamical Bayesian model (Bayesian filter). \textbf{A)} A Bayesian filter consists of the recursive application of two general steps~\cite{kaipo2004}: i) an ~{\em observation update}, which corresponds to novel sensory input and updates the underlying probability distribution of plausible motor commands using the Bayes' formula; ii) a {\em time evolution update}, which denotes the temporal propagation and corresponds to uncertainty increasing with time (see main text); here the probability distribution is updated by convolution with a temporal kernel. These two steps are repeated for each new sensory data in a recursive loop. \textbf{B)} Example distributions for the entire procedure in two scenarios: Gaussian (top) and heavy-tailed (bottom) distributions. The x-axis, $\phi_{t}$, represents the motor command which results in a specific pitch sung by the bird. The outcome of this motor command is then measured by two different sensory modalities, represented by $\{s_{t}^{(i)}\}_{i=1,2}$, with corresponding likelihood functions $L_1(\phi_t; \Delta)$ and $L_2(\phi_t; 0)$, respectively. The $\Delta$ shift for modality 1 is induced by the experimentalist. Dashed brown lines represent the individual likelihood functions from the individual modalities, and the solid lines represent their product, which signals how likely it is that the correct motor command corresponds to $\phi_{t}$. Heavy-tailed distributions can produce a bimodal likelihood, which, multiplied by the prior, suppresses large-error signals. In contrast, Gaussian likelihoods are unimodal and result in greater compensatory changes in behavior.~\label{fig:BayesFilterSketch}}
\end{SCfigure*}

Here we present a theory (Fig.~\ref{fig:BayesFilterSketch}) of a classic model system for sensorimotor learning --- vocal adaptation in a songbird --- in which the brain controls a probability distribution of motor commands and updates this distribution by a recursive Bayesian inference procedure. Since the distribution of the song pitch is empirically heavy tailed (Fig.~\ref{fig:fitting}C), our model does not make the customary Gaussian assumptions. The focus on learning and controlling (non-Gaussian) distributions of behavior allows us to capture successfully all of the above-described nonlinearities in learning dynamics and, furthermore, to account for the previously unnoticed learning-dependent changes in the {\em shape} of the distribution of the behavior.

\section*{Results}
\subsection*{Biological model system}
\begin{figure*}[t!]
\centerline{\includegraphics[width=\textwidth]{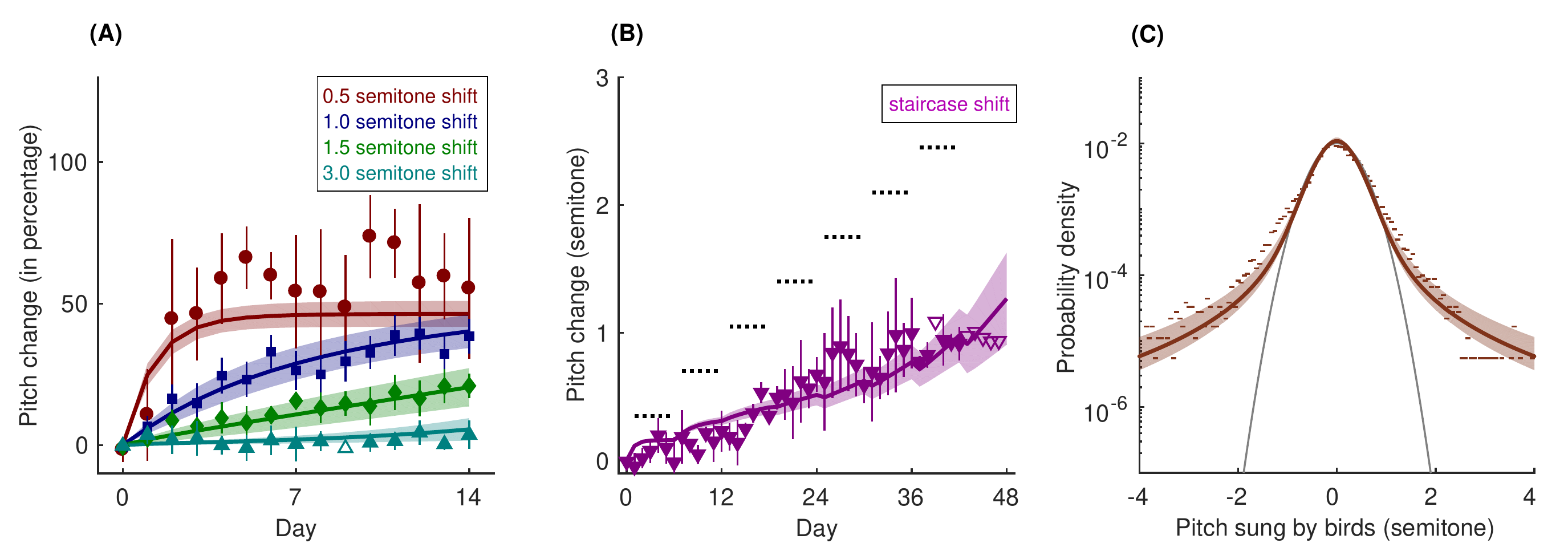}}
\caption{Experimental data and model fitting. The same six parameters of the model are used to {\em simultaneously} fit all data. (A) The dots with error bars are four groups of experimental data, with different colors and symbols indicating different shift sizes (red circle, 0.5 semitone shift; blue square, 1 semitone shift; green diamond, 1.5 semitones shift; cyan upper triangle, 3 semitones shift). The error bars indicate the standard error of the group mean, accounting for variances cross individual birds and within one bird, see {\em Materials and Methods}). For each group, the data are combined from three to eight different birds, and the sign of the experimental perturbation (lowering or raising pitch) is always defined so that adaptive (i.e. error-correcting) vocal changes are positive. Data points without error bars only had a single bird, and they are not used for the fitting, which we denote by hollow symbols. The mean pitch sung on day 0 by each bird is defined as the zero semitone compensation ($\phi=0$). The solid lines with one standard deviation bands (see {\em Materials and Methods}) are results of the model fits, with the same color convention as in experimental data. (B) The dots with error bars show the data from staircase-shift experiment, with the same plotting conventions as in (A). The data are combined from three birds. During the experiment, every six days, the shift size is increased by 0.35 semitone, as shown by the dotted horizontal line segments. On the last day of the experiment, the experienced pitch shift is 2.8 semitones. The magenta solid line with one standard deviation band is the model fit. (C) Dots represent the distribution of pitch on day 0, before the pitch shift perturbation (the baseline distribution), where the data are from 23 different experiments (all pitch shifts combined). The gray parabola is a Gaussian fit to the data within the $\pm1$ semitone range. The empirical distribution has long, non-exponential tails. The brown solid line with one standard deviation band is the model fit.}\label{fig:fitting}
\end{figure*}

Vocal control in songbirds is a powerful model system for examining sensorimotor learning of complex tasks~\cite{brainard2002}. The phenomenology we are trying to explain arises from experimental approaches to inducing song plasticity ~\cite{sober2012}. Songbirds sing spontaneously and prolifically and use auditory feedback to shape their songs towards a ``template'' learned from an adult bird tutor during development. When sensory feedback is perturbed (see below) using headphones to shift the pitch (fundamental frequency) of auditory feedback \cite{sober2012}, birds compensate by changing the pitch of their songs so that the pitch they hear is closer to the unperturbed one.  As shown in Fig.~\ref{fig:fitting}A, the speed of the compensation and its maximum value, measured as a fraction of the pitch shift, decrease with the increasing shift magnitude, so that a shift of 3 semitones results in near-zero fractional compensation.  Crucially, the small compensation for large perturbation does not reflect the limited plasticity of the adult brain since imposing the perturbation gradually, rather than instantaneously, results in a large compensation (Fig.~\ref{fig:fitting}B). 

\subsection*{Data}
We use experimental data collected in our previous work \cite{sober2009,sober2012} to develop our mathematical model of learning. As detailed in Ref.~\cite{kuebrich2015}, we used a virtual auditory feedback system \cite{sober2009,hoffmann2012} to evoke sensorimotor learning in adult songbirds.  For this, miniature headphones were custom-fitted to each bird and used to provide online auditory feedback in which the pitch (fundamental frequency) of the bird's vocalizations could be manipulated in real time, with a loop delay less than 10 ms. In addition to providing pitch-shifted feedback, the headphones blocked the airborne transmission of the bird's song from reaching the ear canals, thereby effectively replacing the bird's natural airborne auditory feedback with the manipulated version. Pitch shifts were introduced after a baseline period of at least 3 days in which birds sang while wearing headphones but without pitch shifts. All pitch shifts were implemented relative to the bird's current vocal pitch and were therefore ``correctable'' in the sense that if the bird changed its vocal pitch to fully compensate for the imposed pitch shift, the pitch of auditory feedback heard through the headphones would be equal to its baseline value. All data were collected during undirected singing (i.e. no female bird was present).

\subsection*{Mathematical model}
To describe the data, we introduce a dynamical Bayesian filter model, Fig.~\ref{fig:BayesFilterSketch}A. We focus on just one variable learned by the animal during repeated singing --- the pitch of the song syllables. Even though the animal learns the motor command and not the pitch directly, we do not distinguish between the produced pitch $\phi$ and the motor command leading to it because the latter is not known in behavioral experiments. We set the mean ``baseline'' pitch sung by the animal as $\phi=0$, representing the ``template'' of song memorized during development, and nonzero values of $\phi$ denote deviations of the sung pitch from the target. 

In our model, the state of the motor learning system at each time step is a probability distribution over motor behaviors, rather than a single motor command or mean behavior that is corrupted by downstream noise as in many other models. Thus at time $t$, the animal has access to the prior distribution over motor commands, $p_\mathrm{prior}(\phi_{t})$. We assume that the bird then randomly selects and produces the pitch from this distribution. In other words, in a departure from standard accounts, we suggest that the experimentally observed variability of sung pitches is dominated by the deliberate exploration of plausible motor commands, rather than by noise in the motor system. This is supported by the experimental finding that the variance of pitch during singing directed at a female (performance) is significantly smaller than the variance during undirected singing (practice)~\cite{kao2005,olveczky2005}. 

After producing a vocalization, the bird then senses the pitch of the produced song syllable through various sensory pathways. Besides the normal airborne auditory feedback reaching the ears, which we can pitch-shift, information about the sung pitch may be available through other, unmanipulated pathways. For example, efference copy may form an internal short term memory of the produced specific motor command ~\cite{niziolek2013}. Additionally, proprioceptive sensing presumably also provides unshifted information~\cite{suthers2002}. Finally, unshifted acoustic vibrations might be transmitted through body tissue in addition to the air, as is thought to be the case in studies that use pitch shifts to perturb human vocal production~\cite{liu2015,scheerer2014}. 

We denote all feedback signals as $s_{t}^{(i)}$ where the index $i$ denotes different sensory modalities. Because sensing is noisy, feedback is not absolutely accurate. Thus it is interpreted using Bayes' formula, so that the posterior probability of which motor commands lead to the target given the observed signals is $p_{\mathrm{post}}(\phi_{t})\propto p_{\mathrm{likelihood}}(\{s_{t}^{(i)}\}|\phi_{t})p_\mathrm{prior}(\phi_{t})$, where $p_{\mathrm{likelihood}}$ represents the distribution of motor commands that would fully ``correct'' the errors given the observed sensory evidence. Finally, the animal expects that the target motor command may change with time because of slow random changes in the motor plant. In other words, the animal must increase its uncertainty about the target with time in the absence of new sensory information. Such increase in uncertainty is given by $p_{\mathrm{prop}}(\phi_{t+\delta t}|\phi_{t})$, the propagator of  statistical field theories \cite{zj}. Overall, this results in the distribution of motor outputs after one cycle of the model
\begin{align}
p_\mathrm{prior}(\phi_{t+\delta t})&=\frac{1}{Z}\int p_\mathrm{prop}(\phi_{t+\delta t}|\phi_{t})\nonumber\\
&\times p_{\mathrm{likelihood}}(\{s_{t}^{i}\}|\phi_{t})p_\mathrm{prior}(\phi_{t})\mathrm{d}\phi_{t},\label{eq:generalBF}
\end{align}
where $Z$ is the normalization constant.

We choose $\delta t$ to be one day in our implementation of the model and lump all vocalizations (which we record) and all sensory feedback (which are unknown) in one time period together. That is, we look at timescales of changes across days, rather than faster fluctuations on timescales of minutes or hours. This matches the temporal dynamics of the learning curves (Fig.~\ref{fig:fitting}A, B). Since the bird sings hundreds of song bouts daily, we now use the law of large numbers and replace the unknown sensory feedback for individual vocalizations by its expectation value $s_{t}^{(i)}\rightarrow \overline{s}_{t}^{(i)}$. For simplicity, we focus on just two sensory modalities, the first affected by the headphones, and the second one not, and we remain agnostic about the exact nature of this second modality among the possibilities noted above. Thus the expectation values of the feedbacks are the shifted and the unshifted versions of the expected value of the sung pitch, $\overline{s}_{t}^{(1)}=\overline{\phi}_{t}-\Delta$ and $\overline{s}_{t}^{(2)}=\overline{\phi}_{t}$. We refer to the conditional probability distribution for each sensory modality as the likelihood functions $L_i(\phi_t)$ for a certain motor command being the target given the observed sensory feedback. Thus assuming that both sensory inputs are independent measurements of the motor output, we rewrite~\eqref{eq:generalBF} as
\begin{align}
p_{\textrm{prior}}(\phi_{t+\delta t})&=\frac{1}{Z}\int p_{\textrm{prop}}(\phi_{t+\delta t}|\phi_{t})\nonumber\\
&\times L_{1}(\phi_{t};\Delta)L_{2}(\phi_{t};0)p_{\textrm{prior}}(\phi_{t})\mathrm{d}\phi_{t},\label{eq:factorizedBF}
\end{align}
where  $0$ and $\Delta$ represent the centers of the likelihoods. 

As illustrated in Fig.~\ref{fig:BayesFilterSketch}B, such Bayesian filtering behaves differently for Gaussian and heavy-tailed likelihoods and propagators. Indeed, if the two likelihoods are Gaussians, their product is also a Gaussian centered between them. In this case, the learning speed of an animal is linear in the error $\Delta$, no matter how large this error is, which conflicts with the experimental results in songbirds and other species~\cite{linkenhoker2002,brainard2002,sober2009,wei2009}. Similarly, if the two likelihoods have long tails, then when the error is small, their product is also a single-peaked distribution as in the Gaussian case. However, when the error size $\Delta$ is large, the product of such long-tailed likelihoods is bimodal, with evidence peaks at the shifted and the unshifted values, with a valley in the middle. Since the prior expectations of the animal are developed before the sensory perturbation is turned on, they peak near the unshifted value. Multiplying the prior by the likelihood then leads to suppression of the shifted peak and hence of large error signals in animal learning. 

In~\eqref{eq:factorizedBF}, there are three distributions to be defined: $L_{1}(\phi_{t};\Delta)$, $L_{2}(\phi_{t};0)$, and $p_{\mathrm{prop}}(\phi_{t+\delta t}|\phi_{t})$, corresponding to the evidence term from the shifted channel, the evidence term from the unshifted channel, and the time propagation kernel, respectively. The prior at the start of the experiment $t=0$, $p_{\mathrm{prior}}(\phi_{0})$, is not an independent degree of freedom: it is the steady state of the recurrent application of  \eqref{eq:factorizedBF} with no perturbation, $\Delta=0$. We have verified numerically that a wide variety of shapes of $L_{1}$, $L_{2}$ and $p_\mathrm{prop}$ result in learning dynamics that can approximate the experimental data (see {\em Materials and Methods}). To constrain the selection of specific functional forms of the distributions, we point out that the error in sensory feedback obtained by the animal is a combination of many noisy processes, including both sensing itself and the neural computation that extracts the pitch from the auditory input and then compares it to the target pitch. By the well-known generalized central limit theorem, the sum of these processes is expected to converge to what are known as Lévy alpha-stable distributions, often simply called stable distribution~\cite{nolan2015} (see {\em Materials and Methods}). If the individual noise sources have finite variances, the stable distribution will be a Gaussian. However, if the individual sources have heavy tails and infinite variances, then their stable distribution will be heavy-tailed as well (Cauchy distribution is one example). Most stable distributions cannot be expressed in a closed form, but they can be evaluated numerically (see {\em Materials and Methods}). Symmetric stable distributions, which we assume here, are characterized by three parameters: the stability parameter $\alpha$ (measuring the proportion of the probability in the tails), the scale or width parameter $\gamma$, and the location or the center parameter $\mu$ (the latter can be predetermined to be $0$, $\Delta$, or the previous time step value in our case). For three distributions $L_{1}(\phi_{t};\Delta)$, $L_{2}(\phi_{t};0)$, and $p_{\mathrm{prop}}$, this results in the total of six unknown parameters.

\subsection*{Fits to data}
We fit the set of six parameters of our model simultaneously to {\em all} the data shown in Fig.~\ref{fig:fitting}. Our dataset consists of twenty-three individual experiments across five experimental conditions: four constant pitch shift learning curves and one gradual, staircase shift learning curve (see {\em Material and Methods} for details). As mentioned previously, birds learn the best (larger and faster compensation) for smaller perturbations, here 0.5-semitone, panel (A). In contrast, for a large 3-semitone perturbation, the birds do not compensate at all within the  14 days of the experiment. However, the birds are able to learn and compensate large perturbations when the perturbation increases gradually, as in the staircase experiment in panel (B). Importantly, the baseline distribution (panel C) has a robust non-Gaussian tail, supporting our model. We note that our six-parameter model fits are able to simultaneously describe {\em all} of these data with a surprising precision, including their most salient features: dependence of the speed and the magnitude of the compensation on the perturbation size for the constant and the staircase experiments, as well as the heavy tails in the baseline distribution.

\begin{figure*}[t!]
\centerline{\includegraphics[width=\textwidth]{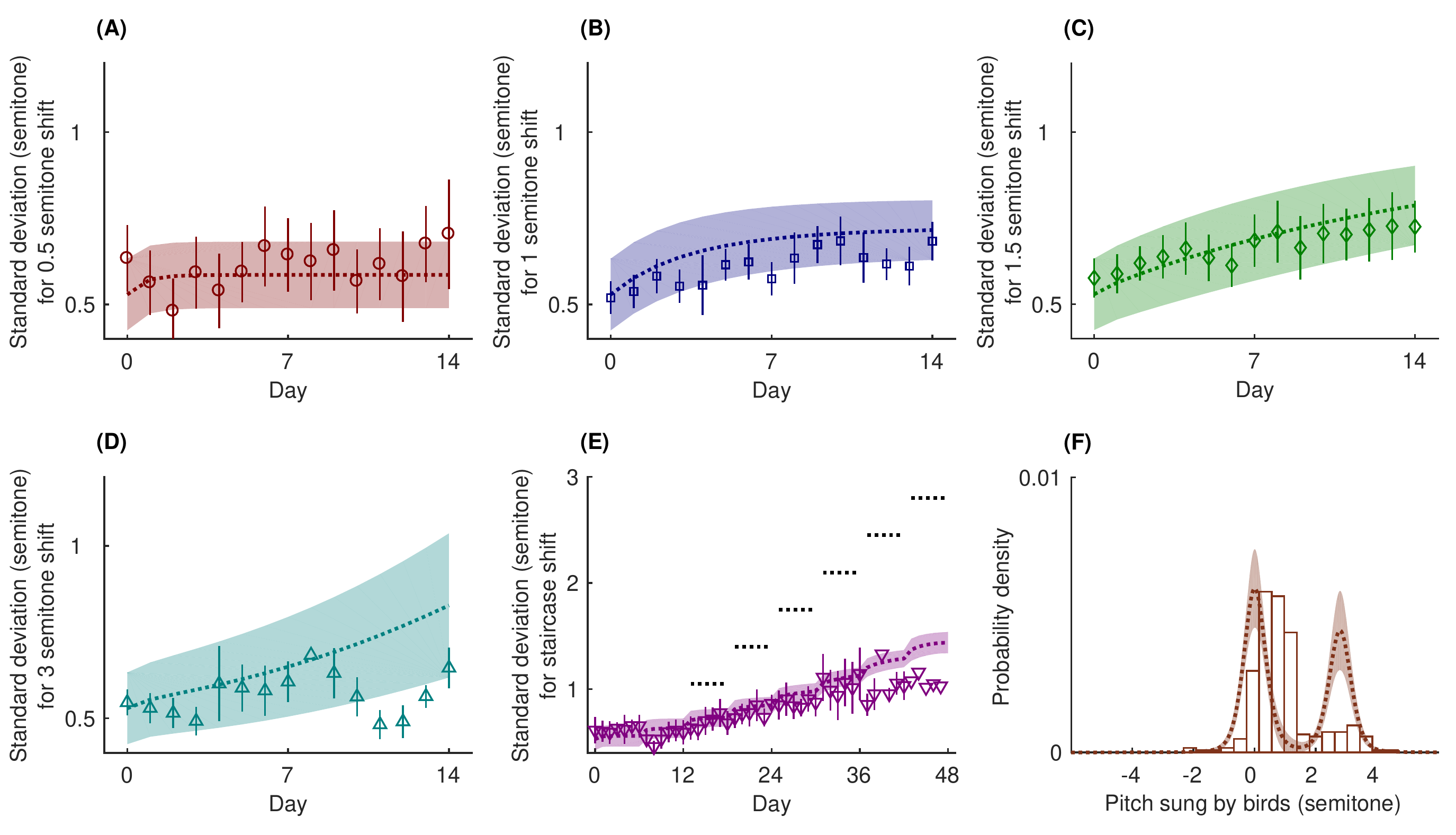}}
\caption{Predictions of our model using the parameter values obtained from fitting the data shown in Fig.~\ref{fig:fitting}. The dots with error bars (A-E) and the histogram (F) represent experimental data with color, symbol, error bar, and other plotting conventions as in Fig.~\ref{fig:fitting}. The dotted lines with one standard deviation bands represent model fits. (A-E) Our model predicts that the standard deviation of the pitch distributions increases gradually in pitch-shift experiments (corresponding to 0.5 semitone, 1 semitone, 1.5 semitones, 3 semitones and staircase shift, respectively). Experiments confirm the prediction. (F) Our model predicts that, at the end of the staircase experiment (mean and standard deviation shown in Figs.~\ref{fig:fitting}B and ~\ref{fig:prediction}E, respectively), the pitch distribution should be bimodal, while it is unimodal initially (cf.~Fig.~\ref{fig:fitting}C), which is also supported by data (note that the data here is from day 47 from the single bird who was exposed to the staircase shift for the longest time).\label{fig:prediction}}
\end{figure*}

\subsection*{Predictions}
Mathematical models are useful to the extent that they can predict experimental results not used to fit them. Quantitative predictions of {\em qualitatively} new results are particularly important for arguing that the model captures the system's behavior. To test the predictive power of our model, we used it to predict the dynamics of higher-order statistics of pitches during learning, rather than using it to simply predict the mean behavior. We first use the model to predict time-dependent measures of the variability (standard deviation in this case) of the pitch. As shown in Figure~\ref{fig:prediction}A-E, our model correctly predicted time-dependent increases in the standard deviation in both single-shift (Figure~\ref{fig:prediction}A-D) and staircase-shift experiments (Figure~\ref{fig:prediction}E) with surprising accuracy. We stress again that no new parameter fits were done for these curves. Potentially even more interestingly, Fig.~\ref{fig:prediction}F shows that our model is capable of predicting unexpected features of the probability distribution of pitches, such as the asymmetric and bimodal structure of the pitch distribution at the end of the staircase-shift experiment. This bimodal structure is predicted by our theory,
since the theory posits that the (bimodal) likelihood distribution (Fig.~\ref{fig:BayesFilterSketch}B, bottom) will iteratively propagate into the observable pitch distribution (the prior). The existence of the bimodal pitch distribution in the data therefore provides strong evidence in support of our theory. Importantly, this phenomenon can never be reproduced by models based on animals learning a single motor command with Gaussian noise around it, rather than a heavy-tailed distribution of motor commands.

\section*{Discussion}

We introduced a novel mathematical framework within the class of observation-evolution models~\cite{kaipo2004} for understanding sensorimotor learning: a dynamical Bayesian filter with non-Gaussian (heavy-tailed) distributions. Our model describes the dynamics of the {\em whole probability distribution} of the motor commands, rather than just its mean value. We posit that this distribution controls the animal's deliberate exploration of plausible motor commands. The model reproduces the learning curves observed in a range of songbird vocal adaptation experiments, which classical behavioral theories have not been able to do to date. Further, also unlike the previous models, our approach predicts learning-dependent changes in the width and shape of the distribution of the produced behaviors.

To further increase the confidence in our model, we show analytically (see {\em Materials and Methods}) that traditional linear models with Gaussian statistics~\cite{kording2007} cannot explain the different levels of compensation for different perturbation sizes.  While we cannot exclude that birds would continue adapting if exposed to perturbations for longer time periods and would ultimately saturate at the same level of adaptation magnitude, the Gaussian models are also argued against by the shape of the pitch distribution, which shows heavy tails (Fig.~\ref{fig:fitting}C and~\ref{fig:prediction}F) and by our ability to predict not just the mean pitch, but the whole pitch distribution dynamics during learning.

An important aspect of our dynamical model is its ability to reproduce multiple different time scales of adaptation (Fig.~\ref{fig:fitting}A, B) using a nonlinear dynamical equation with just a single time scale given by the width of the propagation kernel. As with other key aspects of the model, this phenomenon results from the non-Gaussianity of the distributions employed, and is in contrast to other multiscale models that require explicit incorporation of many time scales~\cite{smith2006,kording2007}. While multiple time scales could be needed to account for other features of the adaptation, our model clearly avoids this for the present data. In the future, we hope that an extension of our model to include multiple explicit time scales will account for individual differences across animals, for the dynamics of acquisition of the song during development, and for the slight shift of the peak of the empirical distribution in Fig.~\ref{fig:prediction}F from $\phi=0$.

Previous analyses of the speed and magnitude of learning in the Bengalese finch have noted that both depend on the overlap of the distribution of the natural variability at the baseline and at the shifted means~\cite{kelly2014,kuebrich2015}: small overlaps result in slower and smaller learning, so that different overlaps lead to different time scales. However, these prior studies have not provided a mechanistic, learning-theoretic explanation of why or how such overlap might determine the dynamics of learning. Our dynamical inference model provides such a mechanism. 

\begin{figure*}[t!]
\centerline{\includegraphics[width=\textwidth]{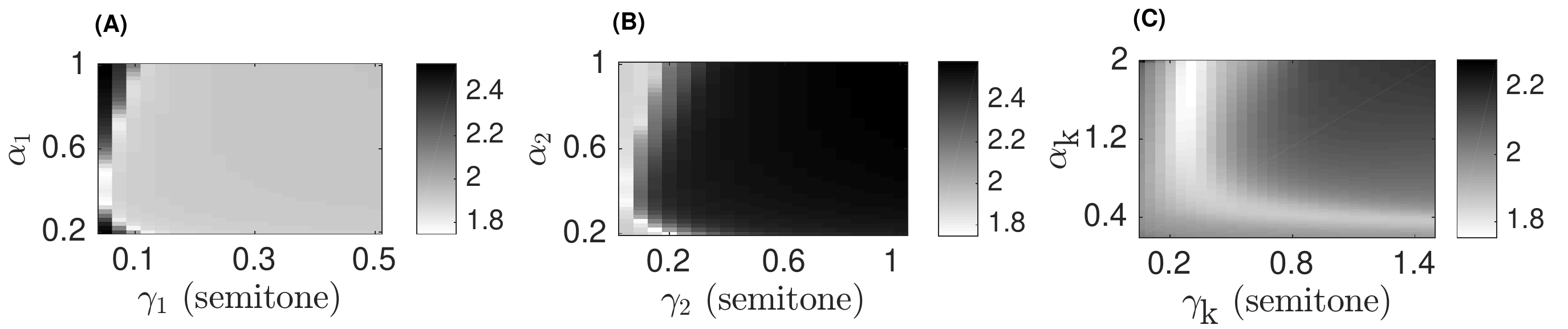}}
\caption{Objective function as a function of two parameters (stability and scale) for (A) the first (shifted) likelihood, (B) the second (unshifted) likelihood, and (C) the propagation kernel, while the respective other four parameters are held fixed. The gray shades represent values in logarithmic scale ($\log_{10}$) of the objective function; lighter color represents a better fit.\label{fig:parameterSpace}}
\end{figure*}

We have chosen the family of so-called Lévy alpha-stable distributions to provide the central ingredient of our model: the heavy tails of the involved probability distributions. In general, a symmetric alpha-stable distribution has a relatively narrow peak in the center and two long fat tails, and this might provide some valuable qualitative insights into how the nervous system processes sensory inputs. For example, a narrow peak in the middle of the likelihood function suggests that the brain puts a high belief in the sensory feedback. However, the heavy tails say that it also puts certain weight (nearly constant) on the probability of very large errors outside of the narrow central region. We have verified that the actual choice of the stable distributions is not crucial for our modeling. For example, one could instead take each likelihood as a power law distribution, or as a sum of two Gaussians with equal means, but different variances. The latter might correspond to mixture of high (narrow Gaussian) and low (wide Gaussian) levels of certainty about sensory feedback, potentially arising from variations in environmental or sensory noise or from variations in attention. As shown in {\em Materials and Methods}, different choices of the underlying distributions result in essentially the same fits and predictions. This suggests that the heavy tails themselves, rather than their detailed shape, are crucial for the model.

While we used Bengalese finches as the subject of this study, nothing in the model relies on the specifics of the songbird system. Sensorimotor learning in many animals should avail itself of modeling by our approach, and we predict that any animal with heavy-tailed distribution of motor outputs should exhibit similar phenomenology in its sensorimotor learning. Exploring whether the model allows for such cross-species generalizations is an important topic for future research, as are questions of how networks of neurons might implement such computations~\cite{fiser2010,buesing2011,kappel2015,petrovici2016}.

\section*{Materials and Methods}

\subsection*{Experiments}
The data used are taken from the experiments in Ref.~\cite{sober2012} and is described in detail there. Briefly, subjects were nine male adult Bengalese finches (females do not produce song) aged over 190 days. Lightweight headphones and microphones were used to shift the perceived pitches of birds' own songs by different amounts, and the pitch of the produced song was recorded. For each day, only data from 10 am to 12 pm is used. The same birds were used in multiple (but not all) pitch shift experiments separated by at least 32 days. Changes in vocal pitch were measured in semitones, which is a relative unit of the fundamental frequency (pitch) of each song syllable:
\begin{align*}
{\textrm{pitch in semitone} \approx 1.2\log_{2} \frac{\textrm{syllable frequency}}{\textrm{mean of baseline syllable frequency}}}.
\end{align*}

\subsection*{Stable distributions}
A probability distribution is said to be stable if a linear combination of two variables distributed according to the distribution has the same distribution up to location and scale~\cite{nolan2015}. By the generalized central limit theorem, the probability distribution of sums of a large number of i.~i.~d.\ random variables with infinite variances tend to be stable distributions~\cite{nolan2015}. A general stable distribution does not have a closed form expression, except for three special cases: Lévy, Cauchy and Gaussian. A symmetric stable variable $x$ can be written in the form $x = \gamma y+\mu$, where $y$ is called the standardized symmetric stable variable and follows the following distribution~\cite{nolan2015}:
\begin{align}
f(y|\alpha)=\frac{1}{2\pi}\int_{-\infty}^{\infty}\mathrm{d}u\ e^{-|u|^{\alpha}}\cos \left(yu\right).
\end{align}
Thus any symmetric stable distribution is characterized by three parameters: the type, or the tail weight, parameter $\alpha$; the scale parameter $\gamma$; and the center $\mu$. $\alpha$ takes the range $(0, 2]$~\cite{nolan2015}. If $\alpha=2$, the corresponding distribution is the Gaussian, and if $\alpha=1$, it is the Cauchy distribution. $\gamma$ can be any positive real number, and $\mu$ can be any real number. The above integral is difficult to compute numerically. However, due to the common occurrence of stable distributions in various fields, such as finance~\cite{mittnik2000}, communication systems~\cite{nikias1995}, and brain imaging~\cite{salas-gonzalez2013}, there are many algorithms to compute it approximately.  We used the method of Ref.~\cite{belov2005}. In this method, the central and tail parts of the distribution are calculated using different algorithms: the central part is approximated by 96-points Laguerre quadrature and the tail part is approximated by Bergstrom expansion~\cite{bergstrom52}. 

Note that even though we take the propagator and the likelihood distributions as stable distributions in our model, their iterative application (effectively, a product of many likelihood distributions) gives finite variance predictions, allowing us to compare predicted variances of the behavior with experimentally measured ones.

\subsection*{Fitting}
Our model consists of three {\em truncated} stable distributions, one for each of the two likelihood functions resembling the feedback modalities and a third for the propagation kernel. We use truncation to ensure biological plausibility: neither extremely large errors nor extremely large pitch changes are physiologically possible. We truncate the distributions to the range $[-8, 8]$ semitones --- much larger than imposed pitch shifts and slightly larger than the largest observed pitch fluctuations in our data, 7 semitones. This leaves us with 9 parameters of which we need to fit 6 from data, namely the type parameters $\alpha$ and the scale parameters $\gamma$, while the center parameters $\mu$ are predetermined: the two likelihoods are at 0 and $\Delta$ respectively, while the propagation kernel is centered around the previous time step value (see~\eqref{eq:generalBF}). We construct an objective function that is a sum of terms representing the quality of fit for the three data sets we consider: the $\chi^2$ for four mean adaptations to the constant shifts (Fig.~\ref{fig:fitting}A), the $\chi^2$ for the mean adaptation for the staircase shift (Fig.~\ref{fig:fitting}B), and the log-likelihood of the observed baseline pitch probability distribution (Fig.~\ref{fig:fitting}C). To make sure that all three terms contribute on about the same scale to the objective function, we multiply the baseline fit term by 10. 

The objective function landscape is not trivial in this case, and there is not a single best set of parameters. Figure~\ref{fig:parameterSpace} illustrates this by showing the quality of fit as a function of each pair of $(\alpha,\gamma)$, while keeping the other four parameters fixed. There is a large subspace (a plateau or a long nonlinear valley, depending on the projection used) that provides nearly the same fit values. In other words, the effective number of important parameters is less than six. Thus choosing the maximum of the objective function and characterizing the error ellipsoid to get the best-fit parameter values and their uncertainties is not appropriate. Instead, we focus on values and uncertainties of the fits themselves. For this, we sweep through the entire parameter space and, for each set of parameters $\vec{\theta}=\{\alpha_{1},\gamma_{1},\alpha_{2},\gamma_{2},\alpha_{k},\gamma_{k}\}$, we calculate the value of the objective function ${\cal L}(\vec{\theta})$ and the corresponding fitted or predicted curve $f(\vec{\theta})$. Then for the mean fits/predictions (lines in our figures), we have
\begin{align}
\left\langle f(\vec{\theta}) \right\rangle=\frac{\sum_{\vec{\theta}}e^{-{\cal L}(\vec{\theta})}f(\vec{\theta})}{\sum_{\vec{\theta}}e^{-{\cal L}(\vec{\theta})}}.\label{eq:objective}
\end{align}
For the standard deviations, denoted by shaded regions in our figures, we have
\begin{align}
\sigma_{f}^{2}=\left\langle f(\vec{\theta})^{2} \right\rangle-\left\langle f(\vec{\theta}) \right\rangle^{2}.\label{eq:sigma}
\end{align}
There are many ways of doing the sweep over the parameters. Here we choose first to find a local minimum (however shallow it is). Then for each parameter, we choose six data points on each side of the minimum, distributed uniformly in the log space between the local minimum and the extremal parameter values ($(0.2,1.9]$ for each $\alpha$ and $[0.01,8]$ for each $\gamma$). The extremal values avoid $\alpha = 0, 2$ and $\gamma=0$, which are singular and dramatically slow down computations. Thus there are total of 13 grid points for each parameter, and the total of $13^6\approx 4.8\cdot 10^6$ total parameter samples. 

\subsection*{Choice of the shape of distributions}
\begin{figure*}[t!]
\centerline{\includegraphics[width=\textwidth]{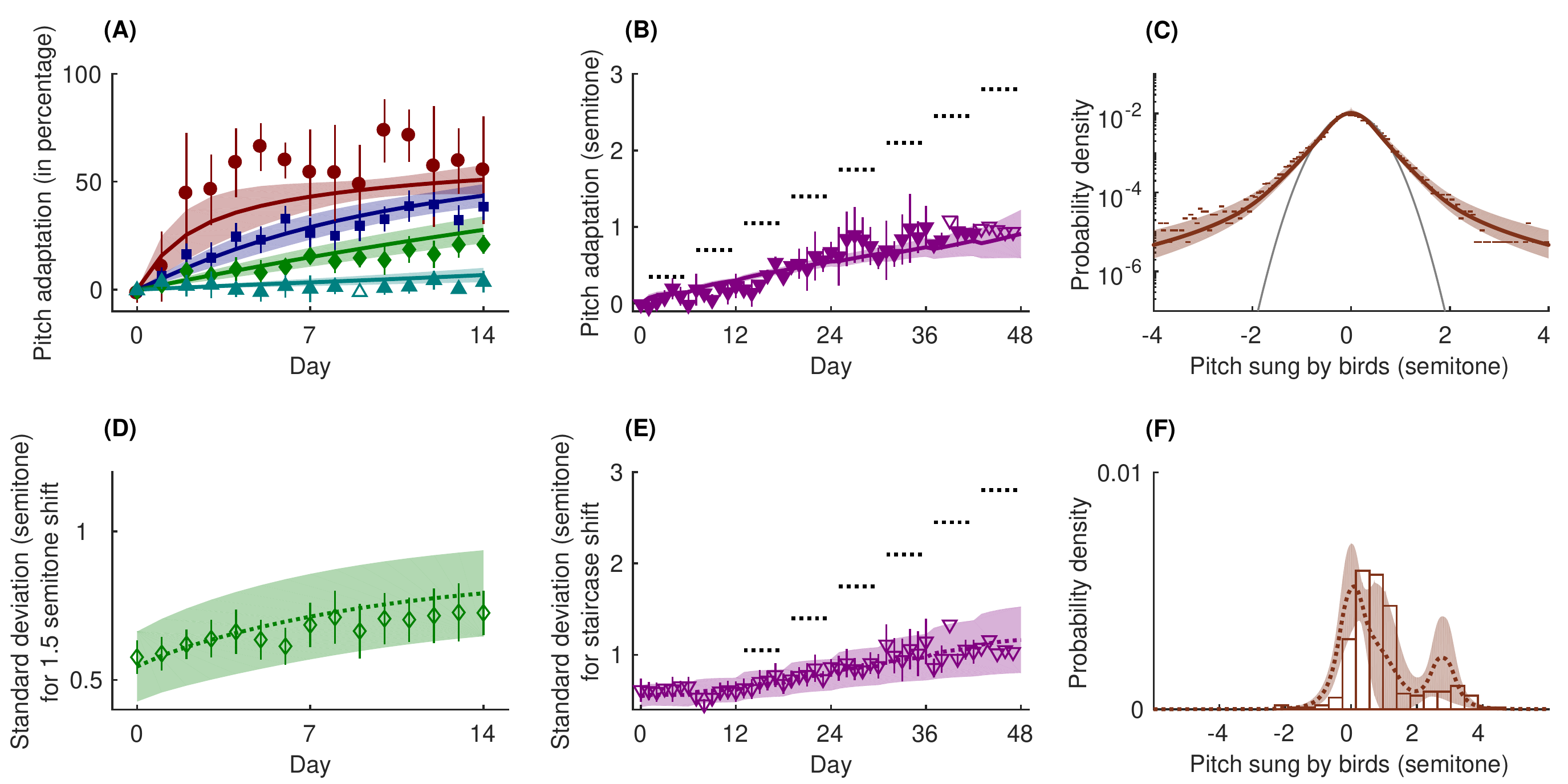}}
\caption{Fits and predictions, same as equivalent panels in Figs.~\ref{fig:fitting}, \ref{fig:prediction}, but with the power law family of heavy tailed distributions instead of the stable distributions family. The shaded areas around the theoretical curves represent confidence intervals for one standard deviation. \label{fig:powerlaw}}
\end{figure*}

For figures in the main text, we have chosen stable distributions for $L_1$, $L_2$ and $p_{\rm prop}$. To investigate effects of this choice, we repeated the fitting and the predictions for different distribution choices. We consider a family of power law distributions $\propto 1/\left(1+(\phi/\gamma)^{2\alpha}\right)$ and a family of mixtures of Gaussians of different width $\rho N(0,\gamma^2)+ (1-\rho)N(0,\delta^2)$.  Distributions in either family produce very similar fits to the stable distribution model. For example, Fig.~\ref{fig:powerlaw} shows the fits and predictions for the power law distribution model.  The detailed shape of the distributions seems less important than the existence of the heavy tails. 

\subsection*{Linear dependence on pitch shift in a Kalman filter with multiple time scales} We emphasized that traditional learning models cannot account for the nonlinear dependence of the speed and the magnitude of learning on the error signal. Here we show this for one such common model, originally proposed by K\"{o}rding et al. ~\cite{kording2007}. This Kalman filter model belongs to the family of Bayes filters, which are dynamical models describing the temporal evolution of the probability distribution of a hidden state variable (can be a vector or a scalar) and its  update using the Bayes formula for integrating information provided by observations, which are conditionally dependent on the current state of the hidden variable. The specific attributes of a Kalman filer within the general class of Bayes filters~\cite{kaipo2004} are the linearity of the temporal evolution of the hidden state (the pitch $\phi$ for the birds, but referred to as {\em disturbances} $d$ in Ref.~\cite{kording2007} and hereon), the linear relation between the measurements (observations) and the hidden variable, and the Gaussian form of the measurement noise and the distribution of disturbances.

One can argue that Kalman filter models with multiple time scales may be able to account for the diversity of learning speeds in our pitch shift experiments. We explore this in the context of an experimentally induced constant shift $\Delta$ to one disturbance $d$ in the Kalman filter model with multiple time scales from Ref.~\cite{kording2007}. If there is a constant shift $\Delta$, Eq.~(3) in Ref.~\cite{kording2007} takes the form
\begin{equation}
O_t = \Delta + \mathrm{H}\cdot\mathrm{d}_t + W_t.
\label{eq:Kalman1}
\end{equation}
The first step in the Kalman filter dynamics is the prediction:
\begin{equation}
\langle \mathrm{d} \rangle_{t+1|t} = \mathrm{A} \langle \mathrm{d} \rangle_{t|t}
\label{eq:Kalman1b},
\end{equation}
where $\langle \mathrm{d} \rangle_{s|t}$ is the mean disturbance vector at time $s$ given measurements up to time $t$ and $\mathrm{A} = \mathrm{diag (1-\tau_i^{-1})}$ with $\tau_i$ being the relaxation timescale of $\mathrm{d_i}$.
We assume that the shift occurs when the disturbances have relaxed to the steady state: $\langle \mathrm{d} \rangle=0$. Therefore, we approximate the standard Kalman filter equation describing the observation update of the expectation value of the disturbance after a measurement at time $t+1$ as (see~\cite{kaipo2004} for a detailed formal description)
\begin{equation}
\langle \mathrm{d} \rangle_{t+1|t+1} =\langle \mathrm{d} \rangle_{t+1|t} + \frac{\Sigma_{t+1|t} \mathrm{H}^T}{\mathrm{H}\Sigma_{t+1|t} \mathrm{H}^T +R}(\Delta - \mathrm{H} \cdot \langle \mathrm{d} \rangle_{t+1|t}),
\label{eq:Kalman2}
\end{equation}
where $R$ is the covariance matrix of the measurement noise, and $\Sigma$ is the covariance matrix of the hidden variables. 
$\Sigma$ does not depend on the measurement and is thus not affected by the  shift $\Delta$. Thus the steady state prediction variance $\Sigma_s$ is given by a solution to the equation
\begin{equation}
\Sigma_s = \mathrm{A} \left( \Sigma_s -\frac{\Sigma_s \mathrm{H}^T\mathrm{H}\Sigma_s}{\mathrm{H}\Sigma_s \mathrm{H}^T +R} \right) \mathrm{A}^T +\mathrm{Q},
\label{eq:Kalman3}
\end{equation}
where $A$ is the matrix determining the temporal evolution of the mean disturbances,~\eqref{eq:Kalman1b}, and $Q$ is the covariance matrix of the intrinsic (temporal evolution) noise.

From~\eqref{eq:Kalman3} we see that $\Sigma_s$ is constant if the perturbation occurs when the system was at the steady state. We now wish to find the new steady-state given the constant perturbation $\Delta$. Consider, for simplicity, two disturbances, each with its own temporal scale $n=2$. The components of the steady state covariance are 
\begin{equation}
\Sigma_s = 
\begin{bmatrix}
    \Sigma_{11} & \Sigma_{12}\\
    \Sigma_{12} & \Sigma_{22}
\end{bmatrix},
\end{equation}
and we define
\begin{align} \nonumber
f_1 &= \frac{\Sigma_{11}+\Sigma_{12}}{\Sigma_{11}+2\Sigma_{12}+\Sigma_{22}},\\
f_2 &= \frac{\Sigma_{12}+\Sigma_{22}}{\Sigma_{11}+2\Sigma_{12}+\Sigma_{22}}.
\label{eqs8}
\end{align}

Substituting~\eqref{eq:Kalman1b} in~\eqref{eq:Kalman2} we get
\begin{eqnarray}
\begin{bmatrix}
\langle d_1 \rangle_{t+1|t+1}\\
\langle d_2 \rangle_{t+1|t+1}
\end{bmatrix}
=
\begin{bmatrix}
    1-\tau_1^{-1} & 0\\
    0 & 1-\tau_2^{-1}
\end{bmatrix}
\begin{bmatrix}
\langle d_1 \rangle_{t|t}\\
\langle d_2 \rangle_{t|t}
\end{bmatrix}
\nonumber
\\
+
\begin{bmatrix}
f_1\\
f_2
\end{bmatrix}
(\Delta - (1-\tau_1^{-1})\langle d_1 \rangle_{t|t} - (1-\tau_2^{-1})\langle d_2 \rangle_{t|t}).
\end{eqnarray}
In the steady state, $\langle \mathrm{d} \rangle_{t+1|t+1}=\langle \mathrm{d} \rangle_{t|t} = \mathrm{d}_s$, we get
\begin{align}
d_s^{(1)} &= \Delta\frac{f_1 \tau_1}{1+f_1(\tau_1-1)+f_2(\tau_2-1)}\\
d_s^{(2)} &= \Delta\frac{f_2 \tau_2}{1+f_1(\tau_1-1)+f_2(\tau_2-1)}
\end{align}
Thus we find that the sum of the disturbances is {\em proportional} to $\Delta$ \emph{independent} of the size of $\Delta$ even for systems with multiple time scales.

Generalizing the result to $n$ disturbances with different time scales, we get the following equations at steady state:
\begin{equation}
\tau_i^{-1} d_s^i = f_i \Delta - f_i\sum_{j=1}^n(1-\tau_j^{-1})d_s^j
\end{equation}
These equations are solved by 
\begin{equation}
d_s^i = \Delta\frac{f_i\tau_i}{1+\sum_{j=1}^n f_j(\tau_j-1)}
\end{equation}
which generalizes the linear dependence of learning on $\Delta$ for arbitrary $n$. Thus this (and similar) Kalman filter based model cannot explain the experimental results studied here.

\begin{acknowledgements}
\noindent This work was partially supported by NIH BRAIN Initiative Theory Grant 1R01-EB022872, James S.\ McDonnell Foundation Grant 220020321, NIH Grant NS084844, and NSF Grant 1456912. We are grateful to the NVIDIA corporation for supporting our research with donated Tesla K40 GPUs.
\end{acknowledgements}

\bibliography{library}

%
%
%
%

\end{document}